\documentclass{llncs}

\usepackage[cp1250]{inputenc}

\usepackage{makeidx}  
\usepackage{amssymb}
\usepackage{amsmath}
\usepackage{epsfig}
\usepackage{subfigure}
\usepackage{bbm}  
\usepackage{calrsfs}

\usepackage{helvet}         
\usepackage{courier}        
\usepackage{type1cm}        
\usepackage{graphicx}        
\usepackage{multicol}        
\usepackage[bottom]{footmisc}

\usepackage{yhmath} 
\usepackage{amsbsy}
\usepackage{amsfonts}
\usepackage{mathbbol}
\usepackage{multirow}
\usepackage{bm}
\usepackage{yfonts}

\usepackage[extra]{tipa}
\usepackage{datetime}  
\usepackage{longtable}
\usepackage[width=0.95\textwidth]{caption}

\newcommand{\olint}{{\overline{\int}}}

\newcommand{\Pprob}{\mathbbm{P}}
\newcommand{\Qprob}{\mathbbm{Q}}

\newtheorem{assumptionnew}[theorem]{Assumption}

\makeindex             

\begin{document}

\allowdisplaybreaks

%
%
\title{
Optimal transport with some directed distances
}
\titlerunning{
Directed optimal transport
}  
%
\author{Wolfgang Stummer 
}
\authorrunning{Stummer} 
%
\tocauthor{Wolfgang Stummer}
\institute{
Department of Mathematics, University of Erlangen--N\"{u}rnberg,
Cauerstrasse $11$,\\
91058 Erlangen, Germany, 
\ \email{stummer@math.fau.de},
\\ 
as well as Affiliated Faculty Member of the School of Business and Economics,
University of Erlangen--N\"{u}rnberg,
Lange Gasse 20, 90403 N\"{u}rnberg, Germany.
}

\maketitle              

\begin{abstract}
We present a toolkit of directed distances
between quantile functions. By employing this,
we solve some new optimal transport (OT) 
problems which e.g. considerably flexibilize 
some prominent OTs expressed through Wasserstein distances.

\end{abstract}



\section{Introduction}
\label{StuGSI21:sec:1}

\vspace{-0.3cm}
Widely used tools for various different tasks in statistics
and probability (and thus, in the strongly connected 
fields of information theory, artificial intelligence and machine learning)
are density-based ``directed'' (i.e. generally asymmetric) distances -- 
called divergences -- 
which measure the dissimilarity/proximity
between two probability distributions; 
some comprehensive overviews  can be found in the books of 
e.g.\ Liese \& Vajda~\cite{StuGSI21:Lie87},
Read \& Cressie~\cite{StuGSI21:Rea88},
Vajda~\cite{StuGSI21:Vaj89}, 
Csisz\'ar \& Shields~\cite{StuGSI21:Csi04},
Stummer~\cite{StuGSI21:Stu04a}, 
Pardo~\cite{StuGSI21:Par06}, 
Liese \& Miescke~\cite{StuGSI21:Lie08},
Basu et al.~\cite{StuGSI21:Bas11}. 
Amongst others, some important 
density-based 
directed-distance classes are:\\  
\noindent
(1) the \textit{Csiszar-Ali-Silvey-Morimoto $\phi-$divergences}
(CASM divergences)~\cite{StuGSI21:Csi63},\cite{StuGSI21:Ali66},\cite{StuGSI21:Mori63}: 
this includes e.g. the total variation distance, exponentials of Renyi cross-entropies,
and the power divergences;
the latter cover e.g. the Kullback-Leibler information divergence (relative entropy), 
the (squared) Hellinger distance, the Pearson chi-square divergence;\\
(2) the \textit{``classical'' Bregman distances} (CB distances)
(see e.g.
Csiszar~\cite{StuGSI21:Csi91},
Pardo \& Vajda~\cite{StuGSI21:Par97,StuGSI21:Par03}): 
this includes e.g. the density-power divergences  
cf.~\cite{StuGSI21:Bas98})
with the squared $L_2-$norm as special case.

\vspace{0.2cm}
\noindent
More generally,
Stummer~\cite{StuGSI21:Stu07b} and
Stummer \& Vajda~\cite{StuGSI21:Stu12a} 
introduced the concept of \textit{scaled Bregman distances}, 
which enlarges and flexibilizes 
both the above-mentioned CASM and CB divergence/distance classes;  
their use for robustness of minimum-distance parameter estimation, testing as well as applications  
can be found e.g. in Ki{\ss}linger \& Stummer~\cite{StuGSI21:Kis13,StuGSI21:Kis15,StuGSI21:Kis16,StuGSI21:Kis18},
Roensch \& Stummer~\cite{StuGSI21:Roe17,StuGSI21:Roe19a,StuGSI21:Roe19b},
Kr{\"o}mer \& Stummer~\cite{StuGSI21:Kro19}.
An even \textit{much wider framework of directed distances/divergences} (BS distances) 
was introduced in the recent comprehensive paper of 
Broniatowski \& Stummer~\cite{StuGSI21:Bro19a}.

\enlargethispage{1.9cm}

\vspace{0.2cm}
\noindent
Another important omnipresent scientific concept is
\textit{optimal transport}; comprehensive general insights into this
(and directly/closely related topics such as e.g. 
mass transportation problems, Wasserstein distances, 
optimal coupling problems, extremal copula problems, 
optimal assignment problems)
can be found in the books of e.g. Rachev \& R\"uschendorf~\cite{StuGSI21:Rac98a},
Villani~\cite{StuGSI21:Vil03,StuGSI21:Vil09}, Santambrogio~\cite{StuGSI21:San15},
Peyre \& Cuturi~\cite{StuGSI21:Pey19},
and the survey paper of e.g. Ambrosio \& Gigli~\cite{StuGSI21:Amb13}. 

\vspace{0.2cm}
\noindent 
In the light of the above explanations,
the main goals of this paper are:
 
\noindent 
(i) To apply (for the sake of brevity, only some subset of) 
the BS distances to the context of \textit{quantile functions}
(see Section 2).

\noindent
(ii) To establish a link between (i)
and a new class of 
optimal transport problems 
where the cost functions are \textit{pointwise} BS distances
(see Section 3).


\vspace{-0.4cm}

\section{A toolkit of divergences between quantile functions}
\label{StuGSI21:sec:2}

\vspace{-0.3cm}
\noindent
By adapting and widening the concept of \textit{scaled Bregman distances} of 
Stummer~\cite{StuGSI21:Stu07b}
and Stummer \& Vajda~\cite{StuGSI21:Stu12a}, 
the recent paper of
Broniatowski \& Stummer \cite{StuGSI21:Bro19a} introduced a
fairly universal, flexible, multi-component system of 
``directed distances'' (which we abbreviate as \textit{BS distances})  
\textit{between two arbitrary functions}; 
in the following, we apply (for the sake of brevity, only parts of) this 
to the important widely used context of quantile functions. 
We employ the following ingredients:

\vspace{-0.6cm}

\subsection{Quantile functions}
\label{StuGSI21:subsec:2.1}
\vspace{-0.2cm}
(I1) Let $\mathcal{X} := ]0,1[$ (open unit interval). 
For two probability distributions $\Pprob$ and $\Qprob$ 
on the one-dimensional Euclidean space $\mathbb{R}$, 
we denote their cumulative distribution functions (cdf) as $F_{\Pprob}$
and $F_{\Qprob}$, and write their quantile functions as

\enlargethispage{0.5cm}

\vspace{-0.7cm}

\begin{eqnarray}
& & \textstyle
F_{\Pprob}^{\leftarrow} 
:= \left\{F_{\Pprob}^{\leftarrow}(x)\right\}_{x \in \mathcal{X}}
:= \left\{\inf\{z \in \mathbb{R}: F_{\Pprob}(z) \geq x \} \right\}_{x \in \mathcal{X}}
\nonumber\\ 
& & \textstyle
F_{\Qprob}^{\leftarrow} 
:= \left\{F_{\Qprob}^{\leftarrow}(x)\right\}_{x \in \mathcal{X}}
:= \left\{\inf\{z \in \mathbb{R}: F_{\Qprob}(z) \geq x \} \right\}_{x \in \mathcal{X}};
\nonumber
\end{eqnarray}

\vspace{-0.3cm}

\noindent
if (say) $\Pprob$ is concentrated on $[0,\infty[$ (i.e., the support of $\Pprob$
is a subset of $[0,\infty[$),
then we alternatively take (for our purposes, without loss of generality)

\vspace{-0.7cm}

\begin{eqnarray}
& & \textstyle
F_{\Pprob}^{\leftarrow} 
:= \left\{F_{\Pprob}^{\leftarrow}(x)\right\}_{x \in \mathcal{X}}
:= \left\{\inf\{z \in [0,\infty[: F_{\Pprob}(z) \geq x \} \right\}_{x \in \mathcal{X}} 
\nonumber
\end{eqnarray} 

\vspace{-0.25cm}

\noindent
which leads to the nonnegativity $F_{\Qprob}^{\leftarrow}(x) \geq 0$ for all $x \in \mathcal{X}$.

Of course, if the underlying cdf $z \rightarrow F_{\Pprob}(z)$ is strictly increasing, then
$x \rightarrow F_{\Pprob}^{\leftarrow}(x)$ is nothing but its ``classical'' inverse function. 
Let us also mention that in quantitative finance and insurance, 
the quantile $F_{\Pprob}^{\leftarrow}(x)$ (e.g. quoted in US dollars units) is called the value-at-risk for confidence level $x \cdot 100\%$. A detailed discussion on properties 
and pitfalls of univariate quantile
functions can be found e.g. in Embrechts \& Hofert~\cite{StuGSI21:Emb13}.

\vspace{-0.6cm}

\subsection{Directed distances --- basic concept}
\label{StuGSI21:subsec:2.2}
\vspace{-0.2cm} 
We quantify 
the \textit{dissimilarity} between the two 
quantile functions $F_{\Pprob}^{\leftarrow}$,$F_{\Qprob}^{\leftarrow}$ 
in terms of \textit{BS distances}
$D^{c}_{\beta}\big(F_{\Pprob}^{\leftarrow},F_{\Qprob}^{\leftarrow}\big)$ 
with $\beta = (\phi,M_{1},M_{2},M_{3})$,
defined by 

\vspace{-0.6cm}

{\footnotesize
\begin{eqnarray} 
& & \hspace{-0.2cm} 
\textstyle
0 \leq D^{c}_{\phi,M_{1},M_{2},M_{3}}\big(F_{\Pprob}^{\leftarrow},F_{\Qprob}^{\leftarrow}\big) 
\nonumber\\ 
& & \hspace{-0.2cm} 
\textstyle
:= \olint_{\negthinspace\negthinspace{\mathcal{X}}}
\Big[ \phi  \Big( {
\frac{F_{\Pprob}^{\leftarrow}(x)}{M_{1}(x)}}\Big) -\phi  \Big( {\frac{F_{\Qprob}^{\leftarrow}(x)}{M_{2}(x)}}\Big)  
- \phi_{+,c}^{\prime} 
\Big( {\frac{F_{\Qprob}^{\leftarrow}(x)}{M_{2}(x)}}\Big) \cdot \Big( \frac{F_{\Pprob}^{\leftarrow}(x)}{M_{1}(x)}-\frac{F_{\Qprob}^{\leftarrow}(x)}{M_{2}(x)}\Big)  
\Big] \cdot M_{3}(x)
\, \mathrm{d}\lambda(x) ; \qquad  
\label{brostu2:fo.def.1}
\end{eqnarray}
}

\vspace{-0.5cm}

\noindent
the meaning of the integral symbol $\olint$ will become clear
in \eqref{brostu2:fo.def.3} below.
Here, in accordance with Broniatowski \& Stummer~\cite{StuGSI21:Bro19a} 
we use:

\begin{itemize}

\vspace{-0.3cm}

\item[(I2)] the Lebesgue measure $\lambda$ 
(it is well known that in general an integral $\int \ldots \mathrm{d}\lambda(x)$
turns --- except for rare cases --- into a classical Riemann integral 
$\int \ldots \mathrm{d}x$).  

\item[(I3)] 
(measurable) \textit{scaling functions}
$M_{1}: \mathcal{X} \rightarrow [-\infty, \infty]$
and 
$M_{2}:\mathcal{X} \rightarrow [-\infty, \infty]$
as well as a
nonnegative 
(measurable) \textit{aggregating function}
$M_{3}: \mathcal{X} \rightarrow [0,\infty]$
such that 
$M_{1}(x) \in ]-\infty, \infty[$, 
$M_{2}(x) \in ]-\infty, \infty[$,
$M_{3}(x) \in [0, \infty[$
for $\lambda$-a.a. $x \in \mathcal{X}$.
In analogy with the above notation, we 
use the symbols $M_{i} := \big\{M_{i}(x)\big\}_{x \in \mathcal{X}}$
to refer to the whole functions.
In the following, $\mathcal{R}\big(G\big)$ denotes the range (image) of 
a function $G := \big\{G(x)\big\}_{x \in \mathcal{X}}$.

\item[(I4)] 
the so-called ``divergence-generator'' $\phi$ 
which is a continuous,
convex (finite) function $\phi: E \rightarrow ]-\infty,\infty[$
on some appropriately chosen open interval $E = ]a,b[$
such that $[a,b]$ covers (at least) the union 
$\mathcal{R}\big(\frac{F_{\Pprob}^{\leftarrow}}{M_{1}}\big) \cup \mathcal{R}\big(\frac{F_{\Qprob}^{\leftarrow}}{M_{2}}\big)$ of both ranges 
$\mathcal{R}\big(\frac{F_{\Pprob}^{\leftarrow}}{M_{1}}\big)$ of 
$\big\{\frac{F_{\Pprob}^{\leftarrow}(x)}{M_{1}(x)}\big\}_{x \in \mathcal{X}}$
and $\mathcal{R}\big(\frac{F_{\Qprob}^{\leftarrow}}{M_{2}}\big)$ of $\big\{\frac{F_{\Qprob}^{\leftarrow}(x)}{M_{2}(x)}\big\}_{x \in \mathcal{X}}$;
for instance, $E=]0,1[$, $E=]0,\infty[$ or $E=]-\infty,\infty[$;
the class of all such functions will be denoted by $\Phi(]a,b[)$.
Furthermore, we assume that $\phi$ is continuously extended 
to $\overline{\phi}: [a,b] \rightarrow [-\infty,\infty]$ by setting 
$\overline{\phi}(t) := \phi(t)$ for $t\in ]a,b[$ as well as  
$\overline{\phi}(a):= \lim_{t\downarrow a} \phi(t)$,
$\overline{\phi}(b):= \lim_{t\uparrow b} \phi(t)$
on the two boundary points $t=a$ and $t=b$. The latter two 
are the only points at which infinite values may appear.
Moreover, for any fixed $c \in [0,1]$
the (finite) function 
$\phi_{+,c}^{\prime}: ]a,b[ \rightarrow ]-\infty,\infty[$
is well-defined by
$\phi_{+,c}^{\prime}(t) := c \cdot \phi_{+}^{\prime}(t)
+ (1- c) \cdot \phi_{-}^{\prime}(t)$, 
where $\phi_{+}^{\prime}(t)$ denotes
the (always finite) right-hand derivative of $\phi$ at the point $t \in ]a,b[$
and $\phi_{-}^{\prime}(t)$ the (always finite) left-hand derivative of $\phi$ at $t \in ]a,b[$. 
If $\phi \in \Phi(]a,b[)$ is
also continuously differentiable -- which we denote by $\phi \in \Phi_{C_{1}}(]a,b[)$ --
then for all $c \in [0,1]$
one gets $\phi_{+,c}^{\prime}(t) = \phi^{\prime}(t)$ ($t \in ]a,b[$)
and in such a situation we always suppress the obsolete indices $c$, $+$ in the corresponding
 expressions.
We also employ the continuous continuation
$\overline{\phi_{+,c}^{\prime}}: [a,b] \rightarrow [-\infty,\infty]$
given by 
$\overline{\phi_{+,c}^{\prime}}(t) := \phi_{+,c}^{\prime}(t)$ ($t \in ]a,b[$), 
$\overline{\phi_{+,c}^{\prime}}(a) := \lim_{t\downarrow a} \phi_{+,c}^{\prime}(t)$, 
$\overline{\phi_{+,c}^{\prime}}(b) := \lim_{t\uparrow b} \phi_{+,c}^{\prime}(t)$.
To explain the precise meaning of \eqref{brostu2:fo.def.1}, we also make 
use of the (finite, nonnegative) 
function
$\psi_{\phi,c}: ]a,b[ \times ]a,b[ \rightarrow [0,\infty[$
given by
$\psi_{\phi,c}(s,t) := \phi(s) - \phi(t) - \phi_{+,c}^{\prime}(t) \cdot (s-t) \geq 0$
($s,t \in ]a,b[$). To extend this to a lower semi-continuous function
$\overline{\psi_{\phi,c}}: [a,b] \times [a,b] \rightarrow [0,\infty]$
we proceed as follows:
firstly, we set $\overline{\psi_{\phi,c}}(s,t) := \psi_{\phi,c}(s,t)$ for all $s,t \in ]a,b[$.
Moreover, since for fixed $t \in ]a,b[$, the function $s \rightarrow \psi_{\phi,c}(s,t)$
is convex and continuous, the limit 
$\overline{\psi_{\phi,c}}(a,t) := \lim_{s \rightarrow a} \psi_{\phi,c}(s,t)$
always exists and (in order to avoid 
overlines in \eqref{brostu2:fo.def.1}) will be 
interpreted/abbreviated as $\phi(a) - \phi(t) - \phi_{+,c}^{\prime}(t) \cdot (a-t)$.
Analogously, for fixed $t \in ]a,b[$ we set 
$\overline{\psi_{\phi,c}}(b,t) := \lim_{s \rightarrow b} \psi_{\phi,c}(s,t)$
with corresponding short-hand notation
$\phi(b) - \phi(t) - \phi_{+,c}^{\prime}(t) \cdot (b-t)$.
Furthermore, for fixed $s\in ]a,b[$ we interpret
$\phi(s) - \phi(a) - \phi_{+,c}^{\prime}(a) \cdot (s-a)$
as 

\vspace{-0.7cm}

\begin{eqnarray} 
& & \hspace{-0.2cm}
\textstyle 
\overline{\psi_{\phi,c}}(s,a) \negthinspace := \negthinspace \big\{
\phi(s) - \overline{\phi_{+,c}^{\prime}}(a) \cdot s
+ \lim_{t \rightarrow a} \big(t \cdot \overline{\phi_{+,c}^{\prime}}(a) - \phi(t) \big) 
\big\}
\cdot \boldsymbol{1}_{]-\infty,\infty[}\big(\overline{\phi_{+,c}^{\prime}}(a)\big)
\nonumber\\[-0.1cm] 
& & \hspace{1.6cm} 
\textstyle 
+ \ \infty \cdot \boldsymbol{1}_{\{-\infty\}}\big(\overline{\phi_{+,c}^{\prime}}(a)\big) \, ,
\nonumber 
\end{eqnarray}

\vspace{-0.3cm}

\noindent
where the involved limit always exists but may be infinite.
Analogously, for fixed $s\in ]a,b[$ we interpret
$\phi(s) - \phi(b) - \phi_{+,c}^{\prime}(b) \cdot (s-b)$
as 

\enlargethispage{0.7cm}

\vspace{-0.5cm}

\begin{eqnarray} 
& & \hspace{-0.2cm} 
\textstyle
\overline{\psi_{\phi,c}}(s,b) := \big\{
\phi(s) - \overline{\phi_{+,c}^{\prime}}(b) \cdot s
+ \lim_{t \rightarrow b} \Big(t \cdot \overline{\phi_{+,c}^{\prime}}(b) - \phi(t) \Big)
\big\}
\cdot \boldsymbol{1}_{]-\infty,\infty[}\big(\overline{\phi_{+,c}^{\prime}}(b)
\big)
\nonumber\\[-0.1cm] 
& & \hspace{1.6cm} 
\textstyle
+ \ \infty \cdot \boldsymbol{1}_{\{+\infty\}}\big(\overline{\phi_{+,c}^{\prime}}(b)\big) \, ,
\nonumber 
\end{eqnarray}

\vspace{-0.2cm}

\noindent
where again the involved limit always exists but may be infinite.
Finally, we always set $\overline{\psi_{\phi,c}}(a,a):= 0$, $\overline{\psi_{\phi,c}}(b,b):=0$,
and $\overline{\psi_{\phi,c}}(a,b) := \lim_{s \rightarrow a} \overline{\psi_{\phi,c}}(s,b)$,
$\overline{\psi_{\phi,c}}(b,a) := \lim_{s \rightarrow b} \overline{\psi_{\phi,c}}(s,a)$.
Notice that $\overline{\psi_{\phi,c}}(\cdot,\cdot)$ is lower semi-continuous but 
not necessarily continuous.
Since ratios are ultimately involved, we also consistently take 
$\overline{\psi_{\phi,c}}\big(\frac{0}{0},\frac{0}{0}\big) := 0$.

\end{itemize}

\vspace{-0.1cm}
\noindent 
With (I1) to (I4),
we define the  
\textit{BS distance (BS divergence)} of \eqref{brostu2:fo.def.1} precisely as

\vspace{-0.6cm}

\begin{eqnarray} 
& & \hspace{-0.2cm} 
\textstyle
0 \leq D^{c}_{\phi,M_{1},M_{2},M_{3}}\big(F_{\Pprob}^{\leftarrow},F_{\Qprob}^{\leftarrow}\big)
= \olint_{\negthinspace\negthinspace{\mathcal{X}}}  
\psi_{\phi,c}\Big(\frac{F_{\Pprob}^{\leftarrow}(x)}{M_{1}(x)},
\frac{F_{\Qprob}^{\leftarrow}(x)}{M_{2}(x)}\Big) \cdot
M_{3}(x) \, \mathrm{d}\lambda(x) \qquad \ 
\label{brostu2:fo.def.2}
\\ 
& & \hspace{-0.2cm} 
\textstyle
: =
\int\displaylimits_{{\mathcal{X}}}  
\overline{\psi_{\phi,c}}\Big(\frac{F_{\Pprob}^{\leftarrow}(x)}{M_{1}(x)},
\frac{F_{\Qprob}^{\leftarrow}(x)}{M_{2}(x)}\Big) \cdot
M_{3}(x) \, \mathrm{d}\lambda(x) ,
\label{brostu2:fo.def.3}
\end{eqnarray}

\vspace{-0.3cm}

\noindent
but mostly use the less clumsy notation with $\olint$ given in \eqref{brostu2:fo.def.1},
\eqref{brostu2:fo.def.2} henceforth,
as a shortcut for the implicitly involved boundary behaviour. 

\vspace{0.15cm}
\noindent
Notice that generally (with some exceptions) one has the asymmetry\\ 
$D^{c}_{\phi,M_{1},M_{2},M_{3}}\big(F_{\Pprob}^{\leftarrow},F_{\Qprob}^{\leftarrow}\big)
\ne 
D^{c}_{\phi,M_{1},M_{2},M_{3}}\big(F_{\Qprob}^{\leftarrow},F_{\Pprob}^{\leftarrow}\big)
$ \\
leading --- together with the nonnegativity ---
to the (already used above) interpretation of
$D^{c}_{\phi,M_{1},M_{2},M_{3}}\big(F_{\Pprob}^{\leftarrow},F_{\Qprob}^{\leftarrow}\big)$
as a \textit{candidate} of a ``directed'' distance/divergence.
To make this \textit{proper}, one needs to verify\\
(NNg) $D^{c}_{\phi,M_{1},M_{2},M_{3}}
\big(F_{\Pprob}^{\leftarrow},F_{\Qprob}^{\leftarrow}\big) \geq 0$.\\[-0.7cm]
\begin{eqnarray} 
& & \hspace{-0.2cm}
\textstyle
\textit{(REg)} \  D^{c}_{\phi,M_{1},M_{2},M_{3}}
\big(F_{\Pprob}^{\leftarrow},F_{\Qprob}^{\leftarrow}\big) = 0 \ \ 
\textrm{if and only if} \ \  
F_{\Pprob}^{\leftarrow}(x) =
F_{\Qprob}^{\leftarrow}(x)
\ \textrm{for $\lambda$-a.a. $x \in \mathcal{X}$.
} 
\qquad \ 
\nonumber 
\end{eqnarray}
As already indicated above, the nonnegativity (NNg) holds per construction.
For the reflexivity (REg) one needs further assumptions.
Indeed, in a more general context beyond quantile functions and the Lebesgue measure, 
Broniatowski \& Stummer~\cite{StuGSI21:Bro19a} gave conditions
such that objects as in \eqref{brostu2:fo.def.2} and \eqref{brostu2:fo.def.3}
satisfy (REg). We shall adapt
this to the current special context, where for the sake of brevity, 
for the rest of this paper we shall always concentrate on the important adaptive subcase 
$M_{1}(x) := W\big(F_{\Pprob}^{\leftarrow}(x),F_{\Qprob}^{\leftarrow}(x) \big)$,
$M_{2}(x) := M_{1}(x)$,
$M_{3}(x) : = W_{3}\big(F_{\Pprob}^{\leftarrow}(x),F_{\Qprob}^{\leftarrow}(x) \big)$,
for some (measurable) functions \\
$W: \mathcal{R}\big(F_{\Pprob}^{\leftarrow}\big) \times \mathcal{R}\big(F_{\Qprob}^{\leftarrow}\big) \rightarrow [-\infty,\infty]$
and
$W_{3}: \mathcal{R}\big(F_{\Pprob}^{\leftarrow}\big) \times \mathcal{R}\big(F_{\Qprob}^{\leftarrow}\big) \rightarrow [0,\infty]$.
Accordingly, \eqref{brostu2:fo.def.1}, \eqref{brostu2:fo.def.2} and
\eqref{brostu2:fo.def.3} simplify to

\vspace{-0.5cm} 

\begin{eqnarray} 
\textstyle
0 & \leq & 
D^{c}_{\phi,W,W_{3}}\big(F_{\Pprob}^{\leftarrow},F_{\Qprob}^{\leftarrow}\big)
:= D^{c}_{\phi,W\big(F_{\Pprob}^{\leftarrow},F_{\Qprob}^{\leftarrow}\big),
W\big(F_{\Pprob}^{\leftarrow},F_{\Qprob}^{\leftarrow}\big),
W_{3}\big(F_{\Pprob}^{\leftarrow},F_{\Qprob}^{\leftarrow}\big)
}\big(F_{\Pprob}^{\leftarrow},F_{\Qprob}^{\leftarrow}\big) 
\nonumber\\ 
\textstyle
& = &
\textstyle
\olint_{\negthinspace\negthinspace{\mathcal{X}}}
\Big[ \phi  \Big( {
\frac{F_{\Pprob}^{\leftarrow}(x)}{W\big(F_{\Pprob}^{\leftarrow}(x),F_{\Qprob}^{\leftarrow}(x) \big)}}\Big) -\phi  \Big( {\frac{F_{\Qprob}^{\leftarrow}(x)}{W\big(F_{\Pprob}^{\leftarrow}(x),F_{\Qprob}^{\leftarrow}(x) \big)}}\Big) 
\nonumber\\
& & \hspace{-0.2cm}  
\textstyle 
- \phi_{+,c}^{\prime} 
\Big( {\frac{F_{\Qprob}^{\leftarrow}(x)}{W\big(F_{\Pprob}^{\leftarrow}(x),F_{\Qprob}^{\leftarrow}(x) \big)}}\Big) \cdot \Big( \frac{F_{\Pprob}^{\leftarrow}(x)}{W\big(F_{\Pprob}^{\leftarrow}(x),F_{\Qprob}^{\leftarrow}(x) \big)}-\frac{F_{\Qprob}^{\leftarrow}(x)}{W\big(F_{\Pprob}^{\leftarrow}(x),F_{\Qprob}^{\leftarrow}(x) \big)}\Big)  
\Big] 
\nonumber\\ 
& & 
\textstyle
\cdot W_{3}\big(F_{\Pprob}^{\leftarrow}(x),F_{\Qprob}^{\leftarrow}(x) \big)
\, \mathrm{d}\lambda(x) \qquad  
\label{brostu2:fo.def.4}
\\
& =: &
\textstyle
\int_{\negthinspace\negthinspace{\mathcal{X}}}  
\overline{\Upsilon}_{\phi,c,W,W_{3}}\big(F_{\Pprob}^{\leftarrow}(x),
F_{\Qprob}^{\leftarrow}(x)\big) \cdot
\, \mathrm{d}\lambda(x) , \qquad \ 
\label{brostu2:fo.def.5}
\end{eqnarray}

\vspace{-0.2cm}

\noindent
where we employ $\overline{\Upsilon}_{\phi,c,W,W_{3}}: 
\mathcal{R}\big(F_{\Pprob}^{\leftarrow}\big) \times \mathcal{R}\big(F_{\Qprob}^{\leftarrow}\big) 
\mapsto [0,\infty]$
defined by

\vspace{-0.5cm}

\begin{eqnarray}
& & \hspace{-0.2cm} 
\textstyle
\overline{\Upsilon}_{\phi,c,W,W_{3}}(u,v) := 
W_{3}(u,v) \cdot
\overline{\psi_{\phi,c}}
\Big(\frac{u}{W\big(u,v\big)},
\frac{v}{W\big(u,v\big)}\Big) \geq 0 \qquad \textrm{with} 
\label{klstzo:fo.def.581}
\\[-0.1cm]
& & \hspace{-0.2cm} 
\textstyle
\psi_{\phi,c}
\Big(\frac{u}{W\big(u,v\big)},
\frac{v}{W\big(u,v\big)}\Big) \negthinspace := \negthinspace
\Big[ \phi \negthinspace \big( {\frac{u}{W(u,v)}}\big) \negthinspace - \negthinspace
\phi \negthinspace \big( {\frac{v}{W(u,v)}}\big) 
\negthinspace - \negthinspace \phi_{+,c}^{\prime} \negthinspace
\big( {\frac{v}{W(u,v)}}\big) \cdot \big( {\frac{u}{W(u,v)}}\negthinspace  - \negthinspace{\frac{v}{W(u,v)}} \big)
\Big]. \quad \ 
\label{klstzo:fo.def.581b}
\end{eqnarray}
We give conditions for the validity of the crucial reflexivity
in the following subsection; this may be skipped by
the non-specialist (and the divergence expert).

\vspace{-0.4cm}

\subsection{Justification of distance properties}
\label{StuGSI21:subsec:2.3}

\vspace{-0.2cm}

\noindent
By construction, one gets for all $\phi \in \Phi(]a,b[)$ and 
all $c \in [0,1]$ the important assertion 
$D^{c}_{\phi,W,W_{3}}\big(F_{\Pprob}^{\leftarrow},F_{\Qprob}^{\leftarrow}\big) \geq 0$
with equality if
$F_{\Pprob}^{\leftarrow}(x) = F_{\Qprob}^{\leftarrow}(x)$
for $\lambda$-almost all $x \in \mathcal{X}$.
As investigated in Broniatowski \& Stummer~\cite{StuGSI21:Bro19a},
in order to get ``sharp identifiability'' (i.e. reflexivity) 
one needs further assumptions on $\phi \in \Phi(]a,b[)$, $c \in [0,1]$;
for instance, if 
$\phi \in \Phi(]a,b[)$ is affine linear on the whole interval $]a,b[$
and $W_{3}$ is constant (say, $1$),
then $\overline{\Upsilon}_{\phi,c,W,W_{3}}$ takes the constant value $0$,
and hence $D^{c}_{\phi,W,W_{3}}\big(F_{\Pprob}^{\leftarrow},F_{\Qprob}^{\leftarrow}\big)= 0$
even in cases where $F_{\Pprob}^{\leftarrow}(x) \ne F_{\Qprob}^{\leftarrow}(x)$ 
for $\lambda$-a.a. $x \in \mathcal{X}$. 
In order to avoid such and similar phenomena, we use the following 

\enlargethispage{0.5cm}

\begin{assumptionnew}
\label{brostu2:assu.class1} 
Let $c \in [0,1]$, $\phi \in \Phi(]a,b[)$ and 

\vspace{-0.2cm}

$$
\textstyle
\mathcal{R}\Big(\frac{F_{\Pprob}^{\leftarrow}}{W\big(F_{\Pprob}^{\leftarrow},F_{\Qprob}^{\leftarrow}\big)}\Big) \,  \cup \,  \mathcal{R}\Big(\frac{F_{\Qprob}^{\leftarrow}}{W\big(F_{\Pprob}^{\leftarrow},F_{\Qprob}^{\leftarrow}\big)}\Big) \, \subset \, [a,b] .
$$ 

\vspace{-0.1cm}

\noindent
Moreover, for all 
$s \in \mathcal{R}\Big(\frac{F_{\Pprob}^{\leftarrow}}{W\big(F_{\Pprob}^{\leftarrow},F_{\Qprob}^{\leftarrow}\big)}\Big)$,   
all 
$t \in \mathcal{R}\Big(\frac{F_{\Qprob}^{\leftarrow}}{W\big(F_{\Pprob}^{\leftarrow},F_{\Qprob}^{\leftarrow}\big)}\Big)$,\\
all $u \in \mathcal{R}\big(F_{\Pprob}^{\leftarrow}\big)$ and
all $v \in \mathcal{R}\big(F_{\Qprob}^{\leftarrow}\big)$
let the following conditions hold:
\begin{itemize}

\vspace{-0.1cm}

\item[(a)] $\phi$ is strictly convex at $t$;

\item[(b)] if $\phi$ is differentiable at $t$ and $s \ne t$, then 
$\phi$ is not affine-linear on the interval $[\min(s,t),\max(s,t)]$
(i.e. between $t$ and $s$); 

\item[(c)] if $\phi$ is not differentiable at 
$t$, $s > t$ and 
$\phi$ is affine linear on $[t,s]$, then we 
exclude $c=1$ for the (``globally/universally chosen'') subderivative
$\phi_{+,c}^{\prime}(\cdot) = c \cdot \phi_{+}^{\prime}(\cdot)
+ (1- c) \cdot \phi_{-}^{\prime}(\cdot)$;

\item[(d)] if $\phi$ is not differentiable at 
 $t$, $s < t$ and 
$\phi$ is affine linear on $[s,t]$, then we exclude $c=0$ for $\phi_{+,c}^{\prime}(\cdot)$;

\item[(e)] $W_{3}\big(u,v\big) < \infty$;

\item[(f)] $W_{3}\big(u,v\big) >0$ if 
$u \ne v$;

\item[(g)] by employing (with a slight abuse of notation)
the function \\
$\Upsilon(u,v):= W_{3}(u,v) \cdot \psi_{\phi,c}
\Big(\frac{u}{W(u,v)},\frac{v}{W(u,v)}\Big)$,
we set by convention\\
$\Upsilon(u,v) := 0$
if $\frac{u}{W(u,v)} = \frac{v}{W(u,v)} = a$;

\item[(h)] 
by convention, $\Upsilon(u,v) := 0$
if $\frac{u}{W(u,v)} = \frac{v}{W(u,v)} = b$;

\item[(i)] 
$\Upsilon(u,v) > 0$ if $\frac{u}{W(u,v)} =a$ and $\frac{v}{W(u,v)} = t \notin \{a,b\}$;\\
this is understood in the following way:
for $\frac{v}{W(u,v)} = t \notin \{a,b\}$, we require\\ 
$\lim_{\frac{u}{W(u,v)} \rightarrow a} \Upsilon(u,v) > 0$
if this limit exists, or otherwise we set by convention 
$\Upsilon(u,v) :=1$ (or any other strictly positive constant) if $\frac{u}{W(u,v)} =a$; 
the following boundary-behaviour conditions have to be
interpreted ana\-logously;

\item[(j)] 
$\Upsilon(u,v) > 0$ if $\frac{u}{W(u,v)} =b$ and $\frac{v}{W(u,v)} = t \notin \{a,b\}$;

\item[(k)] 
$\Upsilon(u,v) > 0$ if $\frac{v}{W(u,v)} =a$ and $\frac{u}{W(u,v)} = s \notin \{a,b\}$;

\item[($\ell$)] 
$\Upsilon(u,v) > 0$ if $\frac{v}{W(u,v)} =b$ and $\frac{u}{W(u,v)} = s \notin \{a,b\}$;

\item[(m)] 
$\Upsilon(u,v) > 0$ if $\frac{v}{W(u,v)} =b$ and $\frac{u}{W(u,v)} = a$\\
(as limit from ($\ell$) or by convention);

\item[(n)] 
$\Upsilon(u,v) > 0$ if $\frac{v}{W(u,v)} =a$ and $\frac{u}{W(u,v)} = b$\\
(as limit from (k) or by convention).

\end{itemize}

\end{assumptionnew}

\enlargethispage{0.2cm}

\vspace{-0.2cm}

\begin{remark}
\label{brostu2:rem.40b}
We could even work with a weaker assumption obtained
by replacing $s$ with  $\frac{F_{\Pprob}^{\leftarrow}(x)}{W\big(F_{\Pprob}^{\leftarrow}(x),F_{\Qprob}^{\leftarrow}(x) \big)}$,
$t$ with $\frac{F_{\Qprob}^{\leftarrow}(x)}{W\big(F_{\Pprob}^{\leftarrow}(x),F_{\Qprob}^{\leftarrow}(x) \big)}$,
$u$ with $F_{\Pprob}^{\leftarrow}(x)$, $v$ with $F_{\Qprob}^{\leftarrow}(x)$,
and by requiring that then the correspondingly plugged-in conditions (a) to (n) 
hold for $\lambda$-a.a. $x \in \mathcal{X}$.
\end{remark}

\noindent
The following requirement is stronger than the 
``model-individual/dependent'' Assumption \ref{brostu2:assu.class1} 
but is more ``universally applicable'':

\vspace{-0.1cm} 

\begin{assumptionnew}
\label{brostu2:assu.class2}
Let $c \in [0,1]$, $\phi \in \Phi(]a,b[)$ on some fixed $]a,b[ \, \in \, ]-\infty,+\infty[$ 

\vspace{-0.3cm}

$$
\textstyle
\textrm{such that } \quad
\mathcal{R}\Big(\frac{F_{\Pprob}^{\leftarrow}}{W\big(F_{\Pprob}^{\leftarrow},F_{\Qprob}^{\leftarrow}\big)}\Big) \,  \cup \,  \mathcal{R}\Big(\frac{F_{\Qprob}^{\leftarrow}}{W\big(F_{\Pprob}^{\leftarrow},F_{\Qprob}^{\leftarrow}\big)}\Big) \, \subset \, ]a,b[ .
$$ 

\vspace{-0.2cm}

\noindent
Moreover, for all 
$s \in \, ]a,b[$, $t \in \, ]a,b[$
all $u \in \mathcal{R}\big(F_{\Pprob}^{\leftarrow}\big)$ and
all $v \in \mathcal{R}\big(F_{\Qprob}^{\leftarrow}\big)$
the conditions (a) to (n) of Assumption \ref{brostu2:assu.class1} hold. 
\end{assumptionnew}

\vspace{-0.1cm}

\noindent
By adapting Theorem 4 and Corollary 1 of Broniatwoski \& Stummer~\cite{StuGSI21:Bro19a},
under Assumption \ref{brostu2:assu.class1} 
(and hence, under Assumption \ref{brostu2:assu.class2}) we obtain\\
(NN) $D^{c}_{\phi,W,W_{3}}\big(F_{\Pprob}^{\leftarrow},F_{\Qprob}^{\leftarrow}\big) \geq 0$.\\[-0.7cm]
\begin{eqnarray} 
& & \hspace{-0.2cm}
\textstyle
\textit{(RE)} \  D^{c}_{\phi,W,W_{3}}\big(F_{\Pprob}^{\leftarrow},F_{\Qprob}^{\leftarrow}\big) = 0 \ \ 
\textrm{if and only if} \ \  
F_{\Pprob}^{\leftarrow}(x) =
F_{\Qprob}^{\leftarrow}(x)
\ \textrm{for $\lambda$-a.a. $x \in \mathcal{X}$.
} 
\qquad \ 
\nonumber 
\end{eqnarray} 

\vspace{-0.3cm}

\noindent
The non-negativity (NN) and the reflexivity (RE) say that
$D^{c}_{\phi,W,W_{3}}\big(F_{\Pprob}^{\leftarrow},F_{\Qprob}^{\leftarrow}\big)$ is indeed
a ``proper'' divergence under Assumption \ref{brostu2:assu.class1}
(and hence, under Assumption \ref{brostu2:assu.class2}).
Thus, the latter will be assumed for the rest of the paper.


\vspace{-0.3cm}

\section{New optimal transport problems}
\label{StuGSI21:sec:3new}

\vspace{-0.2cm}

For our applications to optimal transport,
we impose henceforth the additional requirement that 
the \textit{nonnegative} (extended)
function $\overline{\Upsilon}_{\phi,c,W,W_{3}}$
is continuous and quasi-antitone\footnote{other names are: 
submodular, Lattice-subadditive, 2-antitone, 2-negative, $\Delta-$antitone, supernegative,
``satisfying the (continuous) Monge property/condition''}
in the sense

\vspace{-0.7cm}

\begin{eqnarray}
& & \hspace{-0.2cm} 
\textstyle
\overline{\Upsilon}_{\phi,c,W,W_{3}}(u_{1},v_{1}) + 
\overline{\Upsilon}_{\phi,c,W,W_{3}}(u_{2},v_{2}) 
\leq \overline{\Upsilon}_{\phi,c,W,W_{3}}(u_{2},v_{1}) 
+ \overline{\Upsilon}_{\phi,c,W,W_{3}}(u_{1},v_{2})
\nonumber\\
& & \hspace{6.0cm}
\textstyle
\textrm{for all $u_{1} \leq u_{2}$, $v_{1} \leq v_{2}$}; \qquad \ 
\label{klstzo:fo.def.582}
\end{eqnarray}

\enlargethispage{0.5cm}

\vspace{-0.2cm}

\noindent
in other words, $-\overline{\Upsilon}_{\phi,c,W,W_{3}}(\cdot,\cdot)$
is assumed to be continuous  
and quasi-monotone\footnote{other names are: 
supermodular, Lattice-superadditive, 2-increasing, 2-positive, $\Delta-$monotone, 
2-monotone, 
``fulfilling the moderate growth property'', ``satisfying the measure property'',
``satisfying the twist condition''}
\footnote{
a comprehensive discussion on general quasi-monotone functions can be found e.g. in Chapter 6.C of
Marshall et al.~\cite{StuGSI21:Mar11}}. 
For such a setup, we consider the \textit{novel} Kantorovich transportation problem (KTP) 
with the \textit{pointwise-BS-distance-type} (pBS-type) cost function 
$\overline{\Upsilon}_{\phi,c,W,W_{3}}(u,v)$;
indeed, we obtain the following 

\vspace{-0.2cm}

\begin{theorem}
\label{Stu:thm1}
Let $\widetilde{\Gamma}(\Pprob,\Qprob)$ be
the family of all probability distributions $\mathfrak{P}$ on 
$\mathbb{R} \times \mathbb{R}$
which have marginal distributions $\mathfrak{P}[\, \cdot \times 
\mathbb{R} ] = \Pprob[\cdot]$ 
and $\mathfrak{P}[ \mathbb{R} \times \cdot \, ] = \Qprob[\, \cdot \, ]$. 
Moreover, we denote
the corresponding upper Hoeffding-Fr\'echet bound (cf. e.g. Theorem 3.1.1 of
Rachev \& R\"uschendorf~\cite{StuGSI21:Rac98a}) by
$\mathfrak{P}^{com}$ having ``comonotonic'' 
distribution function $F_{\mathfrak{P}^{com}}(u,v) := \min\{F_{\Pprob}(u),F_{\Qprob}(v)\}$
($u,v \in \mathbb{R}$). Then
\begin{eqnarray}
& & \hspace{-0.2cm} \textstyle
\min_{\{ X \sim \Pprob, \, Y \sim \Qprob \}} \  \mathbb{E}\big[ \    
\overline{\Upsilon}_{\phi,c,W,W_{3}}(X,Y) \, \big] 
\label{klstzo:fo.def.351ot}
\\ 
& & \hspace{-0.2cm} \textstyle 
= \min_{\{ \mathfrak{P} \in \widetilde{\Gamma}(\Pprob,\Qprob) \}} \  
\int\displaylimits_{\mathbb{R} \times \mathbb{R}}
\overline{\Upsilon}_{\phi,c,W,W_{3}}(u,v) \, \mathrm{d}\mathfrak{P}(u,v)
\label{klstzo:fo.def.350ot}
\\ 
& & \hspace{-0.2cm} \textstyle
= \int\displaylimits_{\mathbb{R} \times \mathbb{R}}
\overline{\Upsilon}_{\phi,c,W,W_{3}}(u,v) \, \mathrm{d}\mathfrak{P}^{com}(u,v)
\label{klstzo:fo.def.349ot}
\\ 
& & \hspace{-0.2cm} \textstyle
=  \int_{[0,1]}
\overline{\Upsilon}_{\phi,c,W,W_{3}}(F_{\Pprob}^{\leftarrow}(x),F_{\Qprob}^{\leftarrow}(x)) \, \mathrm{d}\lambda(x)
\label{klstzo:fo.def.348ot}
\\ 
& & \hspace{-0.2cm} \textstyle
= \olint_{]0,1[} \Big[ \phi \negthinspace \left( {
\frac{F_{\Pprob}^{\leftarrow}(x)}{W(F_{\Pprob}^{\leftarrow}(x),F_{\Qprob}^{\leftarrow}(x))}}\right) 
-\phi \negthinspace \left( {\frac{F_{\Qprob}^{\leftarrow}(x)}{W(F_{\Pprob}^{\leftarrow}(x),F_{\Qprob}^{\leftarrow}(x))}}\right)
\nonumber\\ 
& & \hspace{0.2cm} \textstyle 
- \phi_{+,c}^{\prime} \negthinspace
\left( {\frac{F_{\Qprob}^{\leftarrow}(x)}{W(F_{\Pprob}^{\leftarrow}(x),F_{\Qprob}^{\leftarrow}(x))}}\right) \cdot \left( \frac{F_{\Pprob}^{\leftarrow}(x)}{W(F_{\Pprob}^{\leftarrow}(x),F_{\Qprob}^{\leftarrow}(x))}
-\frac{F_{\Qprob}^{\leftarrow}(x)}{W(F_{\Pprob}^{\leftarrow}(x),F_{\Qprob}^{\leftarrow}(x))}\right) 
\Big] \nonumber\\ 
& & \hspace{0.2cm} \textstyle
\cdot
W_{3}(F_{\Pprob}^{\leftarrow}(x),F_{\Qprob}^{\leftarrow}(x)) \, \mathrm{d}\lambda(x) \ 
\label{klstzo:fo.def.347ot}
\\
& & \hspace{-0.2cm} \textstyle
= D^{c}_{\phi,W,W_{3}}\big(F_{\Pprob}^{\leftarrow},F_{\Qprob}^{\leftarrow}\big) 
\geq 0 ,
\label{klstzo:fo.def.XXXot}
\end{eqnarray}

\vspace{-0.2cm}

\noindent
where the minimum in \eqref{klstzo:fo.def.351ot} is taken over
all $\mathbb{R}-$valued
random variables $X$, $Y$ (on an arbitrary probability space $(\Omega, \mathcal{A}, \mathfrak{S})$)
such that $\mathfrak{P}[X \in \cdot \, ] = \Pprob[ \, \cdot \, ]$,
$\mathfrak{P}[Y \in \cdot \, ] = \Qprob[ \, \cdot \, ]$. As usual, 
$\mathbb{E}$ denotes the expectation with respect to $\mathfrak{P}$.
\end{theorem}

\noindent
The assertion \eqref{klstzo:fo.def.349ot}
follows by applying Corollary 2.2a of Tchen~\cite{StuGSI21:Tch80}
(see also -- with different regularity conditions and generalizations --
Cambanis et al.~\cite{StuGSI21:Cam76}, R\"uschendorf~\cite{StuGSI21:Rue83}, Theorem 3.1.2 of Rachev \& R\"uschendorf~\cite{StuGSI21:Rac98a}, 
Theorem 3.8.2 of M\"uller \& Stoyan~\cite{StuGSI21:Mue02}, Theorem 2.5 of Puccetti \& Scarsini~\cite{StuGSI21:Puc10},
Theorem 2.5 of Ambrosio \& Gigli~\cite{StuGSI21:Amb13}).

\begin{remark}
\label{Stu.rem1}
(i) \, Notice that $\mathfrak{P}^{com}$
is $\overline{\Upsilon}_{\phi,c,W,W_{3}}-$independent, and may not be 
the unique minimizer in \eqref{klstzo:fo.def.350ot}. 
As a (not necessarily unique) minimizer
in \eqref{klstzo:fo.def.351ot}, one can take $X := F_{\Pprob}^{\leftarrow}(U)$,
$Y := F_{\Qprob}^{\leftarrow}(U)$ for some 
uniform random variable $U$ on $[0,1]$.\\
(ii) \, In Theorem \ref{Stu:thm1}
we have shown that 
$\mathfrak{P}^{com}$ (cf. \eqref{klstzo:fo.def.349ot}) is an optimal transport plan of
the KTP \eqref{klstzo:fo.def.350ot}
with the \textit{pointwise-BS-distance-type} (pBS-type)
cost function 
$\overline{\Upsilon}_{\phi,c,W,W_{3}}(u,v)$. The outcoming minimal value
is equal to 
$D^{c}_{\phi,W,W_{3}}\big(F_{\Pprob}^{\leftarrow},F_{\Qprob}^{\leftarrow}\big)$
which is typically straightforward to compute (resp. approximate).

\end{remark}

\vspace{-0.1cm}

\noindent
Remark \ref{Stu.rem1}(ii) generally contrasts to those prominently used 
KTP whose cost function is a power
$d(u,v)^{p}$ of 
a metric $d(u,v)$ (denoted as POM-type cost function) which leads to the well-known Wasserstein distances.
(Apart from technicalities) 
There are some overlaps, though:

\vspace{-0.1cm}

\begin{example}
\label{Stu.ex1} 
(i) Take the \textit{non-smooth} $\phi(t) := \phi_{TV}(t):= |t-1|$
($t \in \mathbb{R}$), $c=\frac{1}{2}$,  
$W(u,v) := v$,
$W_{3}(u,v) := |v|$ to obtain
$\overline{\Upsilon}_{\phi_{TV},1/2,W,W_{3}}(u,v) = |u-v| =: d(u,v)$.\\
(ii) Take $\phi(t) := \phi_{2}(t) := \frac{(t-1)^2}{2}$
($t \in \mathbb{R}$, with obsolete $c$),
$W(u,v) :=1$ and $W_{3}(u,v) :=1$ to end up with  
$\overline{\Upsilon}_{\phi_{2},c,W,W_{3}}(u,v) = \frac{(u-v)^2}{2}
= \frac{d(u,v)^2}{2}$.\\
(iii) The \textit{symmetric} distances $d(u,v)$ and $\frac{d(u,v)^2}{2}$ 
are convex functions of $u-v$ and thus continuous quasi-antitone
functions on 
$\mathbb{R} \times \mathbb{R}$.
The correspondingly outcoming Wasserstein distances are thus considerably flexibilized
by our new much more general distance 
$D^{c}_{\phi,W,W_{3}}\big(F_{\Pprob}^{\leftarrow},F_{\Qprob}^{\leftarrow}\big)$
of \eqref{klstzo:fo.def.XXXot}.
\end{example} 

\enlargethispage{0.5cm}

\noindent
Depending on the chosen divergence, one may have to restrict the support
of $\Pprob$ respectively $\Qprob$, for instance to $[0,\infty[$.
We give some further special cases of pBS-type cost functions, 
which are continuous and quasi-antitone, 
but which are generally not symmetric and
thus not of POM-type:

\vspace{-0.1cm}

\begin{example}
\label{Stu.ex2}
``smooth'' pointwise \textit{Csiszar-Ali-Silvey-Morimoto divergences} 
(CASM divergences):
take $\phi :[0,\infty[ \mapsto \mathbb{R}$ to be a strictly convex, twice continuously differentiable
function on $]0,\infty[$ with continuous extension on $t=0$, 
together with $W(u,v) := v$, $W_{3}(u,v) := v$ ($v \in ]0,\infty[$)
and $c$ is obsolete. Accordingly, $\Upsilon_{\phi,c,W,W_{3}}(u,v) := 
v \cdot \phi \negthinspace \left( \frac{u}{v}\right) - v \cdot \phi \negthinspace \left( 1 \right)  
- \phi^{\prime} \negthinspace
\left( 1 \right) \cdot \left( u - v \right) $,
and hence the second mixed derivative satisfies 
$\frac{\partial^{2} \Upsilon_{\phi,c,W,W_{3}}(u,v)}{\partial u \partial v}
= - \frac{u}{v^2} \phi^{\prime\prime} \negthinspace \left( \frac{u}{v}\right) < 0$ 
($u,v \in ]0,\infty[$); thus, $\Upsilon_{\phi,c,W,W_{3}}$ is
quasi-antitone on $]0,\infty[ \times ]0,\infty[$. 
Accordingly,  
\eqref{klstzo:fo.def.351ot} to \eqref{klstzo:fo.def.347ot} 
applies to (such kind of) CASM divergences 
concerning $\Pprob$,$\Qprob$ having support in $[0,\infty[$.
As an example, take e.g. the power function 
$\phi(t):= \frac{t^\gamma-\gamma \cdot t+ \gamma - 1}{\gamma \cdot (\gamma-1)}$
($\gamma \in \mathbb{R}\backslash\{0,1\}$).
A different connection between optimal transport and 
other kind of CASM divergences can be found in Bertrand et al. \cite{StuGSI21:Ber21}
in the current GSI2021 volume.
\end{example}

\begin{example}
\label{Stu.ex3}
``smooth'' pointwise \textit{classical (i.e. unscaled) Bregman divergences} (CBD): 
take $\phi : \mathbb{R} \mapsto \mathbb{R}$ to be a strictly convex, 
twice continuously differentiable function
$W(u,v) := 1$, $W_{3}(u,v) := 1$, 
and $c$ is obsolete. Accordingly, $\Upsilon_{\phi,c,W,W_{3}}(u,v) := 
\phi \negthinspace \left( u\right) -\phi \negthinspace \left( v \right)
- \phi^{\prime} \negthinspace
\left( v \right) \cdot \left( u - v \right) $
and hence $\frac{\partial^{2} \Upsilon_{\phi,c,W,W_{3}}(u,v)}{\partial u \partial v}
= - \phi^{\prime\prime} \negthinspace \left( v\right) < 0$ 
($u,v \in \mathbb{R}$); thus, $\Upsilon_{\phi,c,W,W_{3}}$ is
quasi-antitone on $\mathbb{R} \times \mathbb{R}$.
Accordingly, the representation 
\eqref{klstzo:fo.def.351ot} to \eqref{klstzo:fo.def.347ot} 
applies to (such kind of) CBD. The corresponding special case of
\eqref{klstzo:fo.def.350ot} is called 
``a relaxed Wasserstein distance (parameterized by $\phi$) between
$\Pprob$ and $\Qprob$'' in the recent papers of Lin et al.~\cite{StuGSI21:Lin19} and Guo et al.~\cite{StuGSI21:Guo21}
for a \textit{restrictive} setup where $\Pprob$ and $\Qprob$ are supposed to have 
\textit{compact} support;
the latter two references do 
not give connections to divergences of quantile functions,
but substantially concentrate on applications to topic sparsity
for analyzing user-generated web content and social media, 
respectively, to Generative Adversarial Networks (GANs).
\end{example}

\vspace{-0.4cm}

\begin{example}
\label{Stu.ex4}
``smooth'' pointwise \textit{Scaled Bregman Distances}: for instance, 
consider $\Pprob$ and $\Qprob$ with support in $[0,\infty[$.
Under $W = W_{3}$ one gets that $\Upsilon_{\phi,c,W,W}$ is quasi-antitone on 
$]0,\infty[ \times ]0,\infty[$
if the generator function $\phi$ is strictly convex and
thrice continuously differentiable on $]0,\infty[$ (and hence, c is obsolete) and 
the so-called scale connector $W$ is twice continuously differentiable such that --
on $]0,\infty[ \times ]0,\infty[$ -- 
$\Upsilon_{\phi,c,W,W}$ is twice continuously differentiable 
and $\frac{\partial^2 \Upsilon_{\phi,c,W,W}(u,v)}{\partial u \partial v} \leq 0$
(an explicit formula of the latter is given in the appendix of
Ki{\ss}linger \& Stummer~\cite{StuGSI21:Kis18}). 
Illustrative examples of suitable $\phi$ and $W$ 
can be found e.g. in Ki{\ss}linger \& Stummer \cite{StuGSI21:Kis16}.
\end{example}

\noindent
Returning to the general context,  
it is straightforward to see that if $\Pprob$ does not give mass to points
(i.e. it has continuous distribution function $F_{\Pprob}$) then there exists even a deterministic
optimal transportation plan: indeed, for the map $T^{com} :=  F_{\Qprob}^{\leftarrow} \circ F_{\Pprob}$ 
one has $\mathfrak{P}^{com}[ \, \cdot \, ] = \Pprob[(id,T^{com}) \in \cdot \, ]$
and thus \eqref{klstzo:fo.def.349ot} is equal to

\vspace{-0.7cm}

\begin{eqnarray}
& & \hspace{-0.2cm} \textstyle
\label{klstzo:fo.def.352ot}
\int\displaylimits_{\mathbb{R}} 
\overline{\Upsilon}_{\phi,c,W,W_{3}}(u,T^{com}(u)) \, \mathrm{d}\Pprob(u)
\\  
& & \hspace{-0.2cm} \textstyle
= \min_{\{ T \in \widehat{\Gamma}(\Pprob,\Qprob) \}} \ 
\int\displaylimits_{\mathbb{R}} 
\overline{\Upsilon}_{\phi,c,W,W_{3}}(u,T(u)) \, \mathrm{d}\Pprob(u)
\label{klstzo:fo.def.353ot}
\\ 
& & \hspace{-0.2cm} \textstyle
= \min_{\{ X \sim \Pprob, \, T(X) \sim \Qprob \}} \  \mathbb{E}\big[ \    
\overline{\Upsilon}_{\phi,c,W,W_{3}}(X,T(X)) \, \big]
\label{klstzo:fo.def.354ot}
\end{eqnarray}

\vspace{-0.3cm}

\noindent
where \eqref{klstzo:fo.def.353ot} is called Monge transportation problem (MTP).
Here, $\widehat{\Gamma}(\Pprob,\Qprob)$ denotes the family of all measurable maps
$T: \mathbb{R} \mapsto \mathbb{R}$
such that
$\Pprob[ T \in \cdot \, ] = \Qprob[\, \cdot \, ]$.

\vspace{0.2cm}
\noindent
\textbf{Acknowledgement.} 
I am grateful to the four referees for their comments
and suggestions on readability improvements.

\enlargethispage{0.5cm}

\vspace{-0.4cm}

%
%

\end{document}